\documentclass[prd,twocolumn,showpacs,floatfix,amsmath,amsfonts]{revtex4}

\usepackage{epsfig}
\usepackage{graphicx}
\usepackage{amsmath}
\usepackage{amssymb}
\usepackage{dcolumn}
\begin{document}

\title{High energy density in multi-soliton collisions}

\author{Danial Saadatmand$^{1,2}$}
\email{saadatmand.d@gmail.com}

\author{Sergey V. Dmitriev$^{3,4}$}
\email{dmitriev.sergey.v@gmail.com}

\author{Panayotis G. Kevrekidis$^{5}$}
\email{kevrekid@math.umass.edu }

\affiliation{ $^1$Department of Physics, Ferdowsi University of Mashhad,
91775-1436 Mashhad, Iran
\\
$^2$Department of Physics, University of Sistan and Baluchestan, Zahedan, Iran
\\
$^3$Institute for Metals Superplasticity Problems RAS, Khalturin St.
39, 450001 Ufa, Russia
\\
$^4$National Research Tomsk State University, Lenin Prosp. 36, 634036 Tomsk, Russia
\\
$^5$Department of Mathematics and Statistics, University of
Massachusetts, Amherst, MA 01003 USA }

\begin{abstract}
Solitons are very effective in transporting energy over great distances
and collisions between them can produce high energy density spots
of relevance to phase transformations, energy localization and defect formation among others.
It is then important to study how energy density accumulation scales
in multi-soliton collisions. In this study, we demonstrate
that the maximal energy density that can be achieved in collision of $N$
slowly moving kinks and antikinks in the integrable sine-Gordon field,
remarkably, is proportional to $N^2$, while the total energy of the system is proportional to $N$.
This maximal energy density can be achieved only if the difference between
the number of colliding kinks and antikinks is minimal, i.e., is equal to 0
for even $N$ and 1 for odd $N$ and if the pattern involves an alternating
array of kinks and anti-kinks. Interestingly, for odd (even) $N$ the maximal
energy density appears in the form of potential (kinetic) energy, while kinetic
(potential) energy is equal to zero. The results of the present study rely on
the analysis of the exact multi-soliton solutions for $N=1,2,$ and 3 and on the
numerical simulation results for $N=4,5,6,$ and 7. Based on these results one
can speculate that the soliton collisions in the sine-Gordon field can, in
principle, controllably produce very high energy density. This can have important
consequences for many physical phenomena described by the Klein-Gordon
equations.
\end{abstract}
\maketitle

\section{Introduction}

The celebrated sine-Gordon equation (SGE) \cite{Scott,BookSGE}
\begin{equation}\label{SGE}
\phi_{tt} - \phi_{xx} + \sin \phi = 0,
\end{equation}
has emerged in the geometry of surfaces \cite{Eisenhart} and then it
has long been used in physics to describe propagation of magnetic
flux on an array of superconducting Josephson junctions \cite{Watanabe},
to study the interacting mesons and baryons \cite{Perring}, fermions
in the Thirring model \cite{Coleman}, the properties of crystal dislocations
\cite{BraunKivshar}, dynamics of domain walls in ferromagnetics \cite{Ekomasov}
and ferroelectrics \cite{Ferro1,Ferro2}, the oscillations of an array of pendula \cite{Drazin},
and others \cite{BookSGE,BraunKivshar,Barone,KivshMal}.

The SGE is capable of describing the dynamics of topological solitons
such as a kink and an antikink, as well as their bound state called breather,
a feature that distinguishes it from other continuum models~\cite{weinbirn}. Multi-soliton solutions
to Eq.~(\ref{SGE}) have been derived with the help of the B\"acklund
transformation \cite{Backlund1,Backlund2} or Hirota method \cite{Hirota,Pol}.

However, in addition to its importance in classical mechanics and
also e.g. in condensed matter physics (see e.g.~\cite{giamar} for
a relatively recent example of its use for the description of the
Beresinskii-Kosterlitz-Thouless vortices in superconductors), it is
also an important model in high energy physics. In the latter
context, in addition to its connection to super-symmetric field
theories~\cite{takacs} and string theory~\cite{malda}, it has also
been argued to be related to exotic structures at the interface
of fields and effective particles, such as oscillons~\cite{hindmar}
and Skyrmions (when trapped by vortices)~\cite{skyr}, among others.
Hence, it remains a topic of extensive interest not only within
nonlinear waves but also principally within the theme of fields and
elementary particles.

In the present work, we focus on the energy density arising from
the interaction of prototypical nonlinear structures within the SGE model.
The energy density has a maximum in the kink's core and vanishes away from it.
A moving kink transports this energy as its center of mass moves
and hence kink collisions can result in an increase
of the energy density. For applications it is important to know what is
the largest energy density that can be accumulated in multi-kink collisions.
Such manifestations of large energy density can be associated
with rogue events (i.e., the formation of rogue waves; see e.g.
the reviews of~\cite{pelin,yan}), which are of extreme interest
in recent years. More generally, they can be used for targeted
energy localization which is of interest in its own right.

In this paper, we calculate the maximal energy density that can be achieved
in the collision of $N$ slowly moving sine-Gordon kinks and antikinks for
$N\le 7$. The question is: can the cores of all $N$ colliding solitons
merge at one point, and if yes, what is the maximal energy density
at the collision point? The answers can be readily found in the
concluding Sec.~\ref{Sec:Conclusions} and the way they were obtained
is described in Sec.~\ref{Sec:III}, which follows Sec.~\ref{Sec:II}
with preliminary remarks and a description of the simulation method.
The key result of our considerations is the unexpected scaling of the
maximal energy density (proportional to $N^2$)
with the number of solitons $N$. Furthermore, conditions (on the structure
of the soliton pattern) and manifestations of the energy localization
are illustrated in the process.

\section {Preliminary remarks} \label{Sec:II}

During the dynamics of Eq.~(\ref{SGE}) the total energy
is conserved as:
%
\begin{equation}\label{Energy}
E=K+P,
\end{equation}
which is the sum of the kinetic and potential energies given, respectively, by
\begin{equation}\label{KPEnergy}
K=\int\limits_{-\infty}^{\infty} \frac{1}{2} \phi_t^2 dx, \quad
P=\int\limits_{-\infty}^{\infty} \Big(\frac{1}{2}\phi_x^2 +1-\cos\phi \Big)dx.
\end{equation}

The kinetic energy density and the potential energy
density of the SGE field are given
by the integrands of Eq.~(\ref{KPEnergy}),
\begin{equation}\label{KPEdensity}
    k(x,t)= \frac{1}{2}\phi_t^2, \quad
    p(x,t)= \frac{1}{2}\phi_x^2 +1 - \cos \phi,
\end{equation}
and the total energy density is
\begin{equation}\label{Edensity}
    e(x,t)=k(x,t)+p(x,t).
\end{equation}

\begin{figure}
\includegraphics[width=9cm]{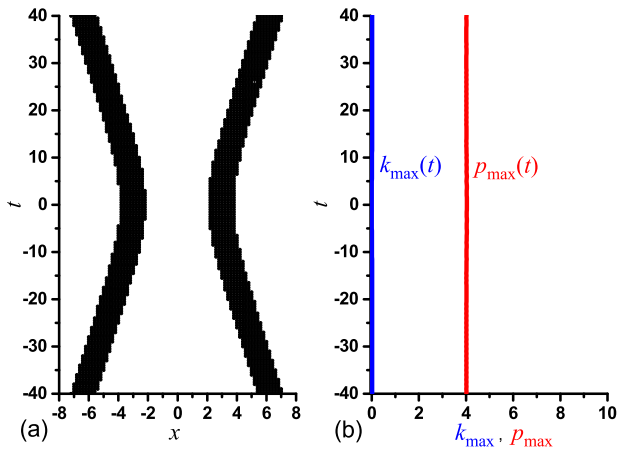}
\caption{Collision of two kinks having velocities $V_k=\pm 0.1$
according to Eq.~(\ref{KK}). (a) Trajectories of the
soliton cores are shown by the regions where total energy density
$e(x,t)>2$. (b) Maximal over spatial coordinate kinetic (blue)
and potential (red) energy densities as the functions of time.
Maximal energy density does not grow during collision of solitons
having the same topological charge because they repel each other
and their cores do not merge.}\label{fig1}
\end{figure}

\begin{figure}
\includegraphics[width=9cm]{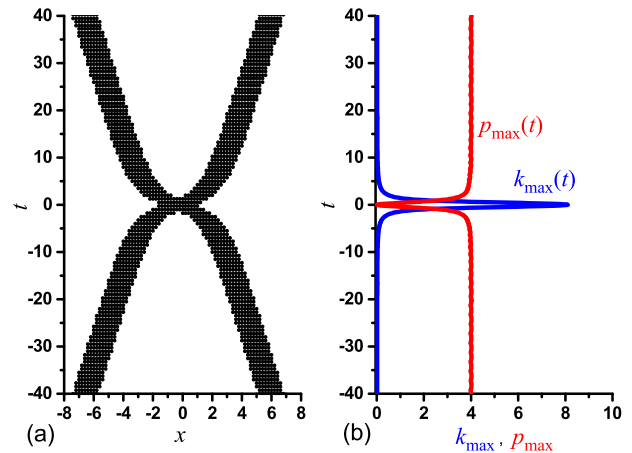}
\caption{Same as in Fig.~\ref{fig1} but for the collision of a kink
and an antikink given by Eq.~(\ref{KA}) with subkink velocities
$V_k=\pm 0.1$. Cores of the mutually attractive solitons merge
at the collision point and total energy density at the collision
point $e(x,0)=k(x,0)+p(x,0)$ rises up to about 8.}\label{fig2}
\end{figure}

\begin{figure}
\includegraphics[width=9cm]{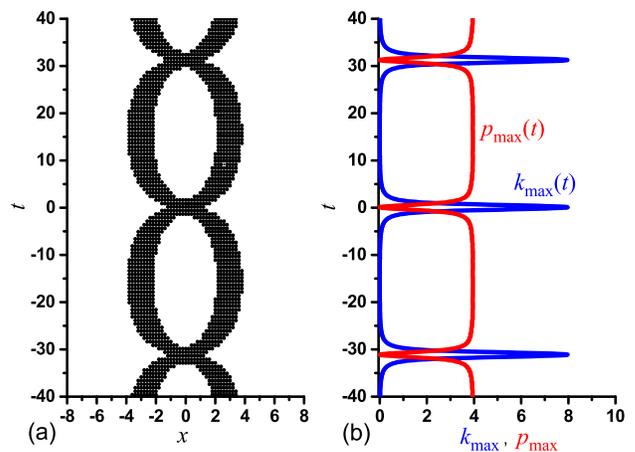}
\caption{Same as in Fig.~\ref{fig1} but for the breather given by
Eq.~(\ref{Breather}) with $V_b=0$ and $\omega=0.1$. When the subkinks
collide, the potential energy density is almost zero and the kinetic
energy density is about 8.}\label{fig3}
\end{figure}

The two basic soliton solutions to SGE (\ref{SGE}) are
the kink (antikink)
\begin{equation}\label{Kink}
   \phi(x,t)=\pm 4\arctan\{\exp[\delta_k(x-V_k t)]\},
\end{equation}
and the breather
\begin{equation}\label{Breather}
   \phi(x,t)=4\arctan\frac
   {\eta\sin[\delta_b\omega(t-V_b x)]}
   {\omega\cosh[\delta_b\eta(x-V_b t)]},
\end{equation}
where $V_k$ is kink velocity, $V_b$, $\omega$ are the breather
velocity and frequency, and
\begin{equation}\label{deltaeta}
   \delta_{k,b} =\frac{1}{ \sqrt{1 - V_{k,b}^2}},\quad \eta =
   \sqrt{1-\omega^2}.
\end{equation}
The upper (lower) sign in Eq.~(\ref{Kink}) corresponds to the kink
(antikink). The breather solution Eq.~(\ref{Breather}) can be regarded as a
kink-antikink bound state \cite{Legrand,Legrand1,CaputoFlytzanis}.

A collision between two kinks having velocities $\pm V_k$ is
described by the following solution to Eq.~(\ref{SGE})
\begin{equation}\label{KK}
   \phi(x,t)=4\arctan\frac{V_k\sinh(\delta_k x)}{\cosh(\delta_k V_k t)}.
\end{equation}

For the collision between kink and antikink having velocities $\pm
V_k$ one has the exact solution
\begin{equation}\label{KA}
   \phi(x,t)=4\arctan\frac{V_k\cosh(\delta_k x)}{\sinh(\delta_k V_k t)}.
\end{equation}

Substituting Eq.~(\ref{Kink}) and Eq.~(\ref{Breather}) into Eq.~(\ref{Energy})
one finds the total energies of the kink and breather
\begin{equation}\label{KBEnergy}
   E_k=8\delta_k, \quad E_b=16\delta_k\eta.
\end{equation}

We are not interested in the relativistic effects and only slow solitons
($V_k \ll 1$, $V_b \ll 1$) will be considered so that $\delta_k\approx 1$
and $\delta_b\approx 1$. Only low-frequency breathers ($\omega \ll 1$)
will be discussed so that $\eta\approx 1$. Then, we
can write approximately that $E_k\approx 8$ and
$E_b\approx 16$.

Even though the analytical expressions for multi-soliton solutions to SGE
are available \cite{Backlund1,Backlund2,Hirota} their complexity increases
rapidly with the number of solitons, $N$. That is why for $N\ge 4$ we will
do calculations numerically. For this we discretize Eq.~(\ref{SGE}) as
follows
\begin{eqnarray}\label{SGEdiscrete}
\frac{d^2\phi_n}{dt^2} - \frac{1}{h^2}(\phi_{n-1} -2\phi_{n}+\phi_{n+1}) \nonumber \\
+\frac{1}{12h^2}(\phi_{n-2}-4\phi_{n-1} +6\phi_{n}-4\phi_{n+1}+\phi_{n+2})\nonumber \\
+\sin\phi_n  = 0,
\end{eqnarray}
where $h$ is the lattice spacing, $n=0,\pm1,\pm2,...$, and $\phi_n(t)=\phi(nh,t)$.
To minimize the effect of discreteness, the term $\phi_{xx}$ in Eq.~(\ref{SGE})
is discretized with the accuracy $O(h^4)$, which has been used previously
\cite{BraunKivshar,KivshMal,Danial}. The equations of motion
in the form of Eq.~(\ref{SGEdiscrete}) were integrated with respect to the time using
an explicit scheme with the time step $\tau$ and the accuracy of $O(\tau^4)$.
The simulations reported in Sec.~\ref{Sec:III} were carried out for $h = 0.1$,
$h = 0.05$ and $\tau = 0.005$.

Before we start the presentation of the main results the following remark
should be made. Two kinks (or two antikinks) repel each other as quasi-particles having the same
topological charge. When they collide, they bounce off each other, their
cores do not merge and, consequently, the maximal energy density does not
grow. This is illustrated by Fig.~\ref{fig1}, where for the solution Eq.~(\ref{KK})
with $V_k=0.1$ we show (a) the regions of the $(x,t)$ plane where the total
energy density $e(x,t)>2$ and (b) the maximal over $x$ densities of kinetic
(blue line) and potential (red line) energies. The kinks collide at $t=0$,
$x=0$. One can see that $k_{\max}(t)$ is nearly zero (due to the small
kink velocity and the quadratic dependence on it), while $p_{\max}(t)\approx 4$,
and these values are not affected by the collision.

On the contrary, kink and antikink are mutually attractive quasi-particles.
Their cores merge during collision and the maximal energy density increases
at the collision point. This can be seen in Fig.~\ref{fig2} where the
kink-antikink solution Eq.~(\ref{KA}) is presented for $V_k=0.1$.
Far from the collision ($t=0$, $x=0$) we have $k_{\max}(t)\approx 0$ and
$p_{\max}(t)\approx 4$. However, at $t=0$ $p_{\max}$ drops to zero, while
$k_{\max}$ rises up to nearly 8, and so does the maximal total energy density
$e_{\max}$ (not shown in the figure).

In Fig.~\ref{fig3} similar results are shown for the breather solution
Eq.~(\ref{Breather}) with $V_b=0$ and $\omega=0.1$. It was already mentioned
that the low-frequency breather can be envisioned as a kink-antikink bound state, and when
the sub-kinks collide, $p_{\max}$ drops to zero and $k_{\max}$ reaches the
value of nearly 8, as does $e_{\max}$. In the breather case, instead
of this happening once (as in Fig.~\ref{fig2}) the phenomenology
periodically repeats itself, due to the time-periodicity of the state.

For the three-kink solutions it has been demonstrated that the cores of all
three kinks can merge only if they collide in the spatial arrangement kink-antikink-kink
(or antikink-kink-antikink) \cite{PREcollisions}. This is understandable because
in the combinations such as kink-kink-kink or kink-kink-antikink the solitons
having the same topological charge repel each other because between them there is no
a soliton of the opposite charge. In the following we will consider the
multi-soliton solutions with alternating kinks and antikinks.
In this case each kink (or antikink) attracts the nearest neighbors of the opposite charge
and all of them can collide at one point, as it will be demonstrated in the
following Section. This type of configurations promotes the energy exchange,
contrary to what is the case for configurations bearing adjacent waves
of the same type.

\section {Maximal energy density of multi-soliton solutions to the SGE} \label{Sec:III}

\subsection{Case $N=1$} \label{Sec:N1}

For a standing kink, Eq.~(\ref{Kink}) with $V_k=0$, the kinetic energy is zero.
Then, the total energy density can be found from
Eqs.~(\ref{KPEdensity})--(\ref{Edensity}) in the form
\begin{equation}\label{K1}
   e(x)=p(x)=8\left(\frac{e^x}{1+e^{2x}} \right)^2 +1-\cos(4\arctan e^x).
\end{equation}
This function has maximum at $x=0$, which is the coordinate of the kink's center.
The value of the maximal energy density of the standing kink is
\begin{equation}\label{K1max}
   e^{(1)}_{\max}=p^{(1)}_{\max}=4.
\end{equation}

\subsection{Case $N=2$} \label{Sec:N2}

\begin{figure}
\includegraphics[width=9cm]{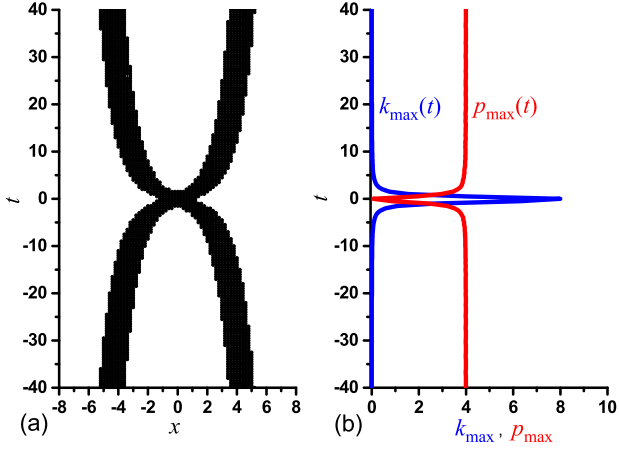}
\caption{Same as in Fig.~\ref{fig1} but for the separatrix two-soliton
solution of Eq.~(\ref{Separt2}). When the kink and antikink collide the
potential energy density is equal to zero and kinetic
energy density is exactly 8.}\label{fig4}
\end{figure}

The breather solution (\ref{Breather}) for $V_b=0$ in the
limit $\omega \rightarrow 0$, and the kink-antikink solution
(\ref{KA}) in the limit $V_k \rightarrow 0$ both approach the
same separatrix two-soliton solution
\begin{equation}\label{Separt2}
   \phi(x,t)=4\arctan\frac{t}{\cosh x}.
\end{equation}
This solution describes the kink and antikink that after the collision
at $t=0$ move apart and their velocities vanish as $t\rightarrow \infty$.
The solution is depicted in Fig.~\ref{fig4} where, as before, in (a) the
points of the $(x,t)$ plane with $e(x,t)>2$ are shown and in (b) the maximal
-- over the spatial coordinate $x$ --
values of kinetic (blue) and potential (red) energy
densities are presented as functions of time. We now calculate the
exact value of the maximal energy density by substituting Eq.~(\ref{Separt2})
into Eqs.~(\ref{KPEdensity})--(\ref{Edensity}). The calculation can be
simplified by noting that at $t=0$ one has $\phi(x,0)\equiv 0$ and thus,
at the collision point the energy of the kink-antikink pair is in the form
of kinetic energy,
\begin{equation}\label{K2}
   e(x,0)=k(x,0)=\frac{8}{\cosh^2 x}.
\end{equation}
The energy density has maximum at the collision point $x=0$:
\begin{equation}\label{K2max}
   e^{(2)}_{\max}=k^{(2)}_{\max}=8.
\end{equation}

\subsection{Case $N=3$} \label{Sec:N3}

\begin{figure}
\includegraphics[width=9cm]{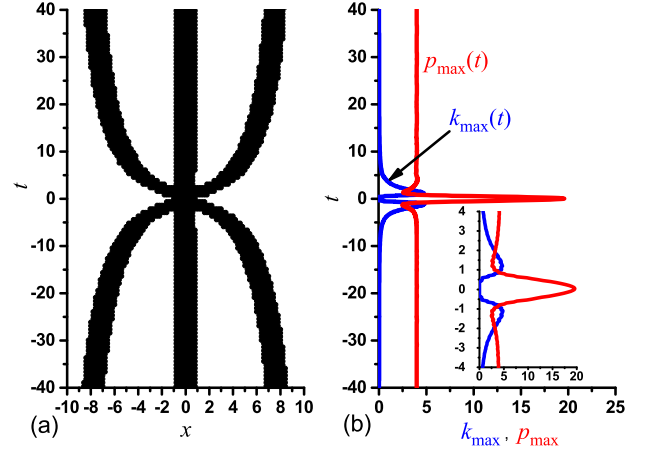}
\caption{Same as in Fig.~\ref{fig1} but for the separatrix three-soliton
solution Eq.~(\ref{KBS}). The inset shows the blowup of the region around
$t=0$. When the two kinks collide with the antikink, the kinetic energy
density is equal to zero and the potential energy density is exactly 20.}
\label{fig5}
\end{figure}

The kink-breather (3-soliton) solution to SGE reads
\begin{eqnarray}\label{KB}
&& \phi(x,t)=4\arctan(\exp B)+4\arctan\frac{\eta Y}{\omega Z}, \nonumber\\
&&Y=2\omega (\sinh D-\cos C \sinh B)\nonumber\\
&&+2\delta_b  \delta_k (V_k -V_b ) \sin C \cosh B,\nonumber \\
&&Z=2\eta (\cos C +\sinh D \sinh B)\nonumber\\
&&-2\delta_b \delta_k (1-V_k V_b ) \cosh D \cosh B,\nonumber \\
&&B=\delta_k (x-V_k t),\nonumber\\
&&C=-\omega\delta_b (t-V_b x), \quad D=\eta\delta_b (x-V_b t).
\end{eqnarray}
In the limit $V_k\rightarrow 0$, $V_b\rightarrow 0$, and $\omega\rightarrow 0$,
this solution assumes the following form (see Eq.~(26) of Ref.~\cite{Mirosh})
\begin{equation}\label{KBS}
\phi(x,t)=4\arctan e^x+4\arctan\frac{x\cosh x-t^2\sinh x}{t^2+\cosh^2x}.
\end{equation}
This separatrix solution describes the antikink standing at $x=0$ and two kinks
that after the collision with the antikink at $t=0$ move apart and
their velocities vanish as $t\rightarrow 0$. The solution is presented
in Fig.~\ref{fig5} using a visualization similar to the previous figures.

To calculate the exact value of the maximal energy density,  we
again substitute Eq.~(\ref{KBS}) into Eqs.~(\ref{KPEdensity})--(\ref{Edensity}). Note that at
$t=0$ one has $\phi_t(x,0)\equiv 0$ and thus, at the collision point the
energy of the kink-antikink-kink solution is in the form of potential energy,
\begin{eqnarray}\label{K3}
   e(x,0)=p(x,0)=\nonumber \\
   8\left(\frac{e^x}{1+e^{2x}}+\frac{\cosh x-x\sinh x}{\cosh^2x+x^2} \right)^2\nonumber \\
   +1-\cos\left(4\arctan e^x +4\arctan\frac{x}{\cosh x}\right).
\end{eqnarray}
The energy density has maximum at the collision point $x=0$:
\begin{equation}\label{K3max}
   e^{(3)}_{\max}=p^{(3)}_{\max}=20.
\end{equation}

\subsection{Case $N=4$} \label{Sec:N4}

The solution to SGE that describes collision of two breathers
(i.e., a 4-soliton solution) with
velocities $V_1$, $V_2$ and frequencies $\omega_1$, $\omega_2$ is given by
\begin{widetext}
\begin{align}\label{BB}
   \phi(x,t)=4\arctan(S)-4\arctan\frac{\eta_2 (T\cosh B_1+\sin C_1)}
   {\omega_2 (\cosh B_1+T\sin C_1)}, \nonumber \\
 T=\varphi\frac{2\tau[(S-P)(1+SP)-Q^2S]-2\beta Q(1+S^2)}
  {\varphi^2[(1+S)(1+SP)+Q^2S^2]+(\tau^2+\beta^2)[(S-P)^2+Q^2]}, \nonumber \\
   P=\frac{\beta X+\kappa Y}{\varepsilon Z}, \quad
    Q=\frac{\beta Y-\kappa X}{\varepsilon Z}, \quad
    S=\frac{\eta_1 \sin C_1 }{\cosh B_1\omega_1}, \nonumber \\
    X=\sinh B_2\cos C_1-\cos C_2\sinh B_1, \,\,
    Y=\cosh B_2\sin C_1+\sin C_2\cosh B_1, \nonumber \\
    Z=\cos(C_1-C_2)+\cosh(B_1+B_2), \nonumber \\
    B_{1,2}=\eta_{1,2} \delta_{1,2}(x-x_{1,2}-V_{1,2}t), \,\,
    C_{1,2}=\Delta_{1,2}-\omega_{1,2} \delta_{1,2}[t-(x-x_{1,2})V_{1,2}], \nonumber \\
    \delta _{1,2}=(1-V_{1,2}^2)^{-1/2}, \,\,
    \eta_{1,2}=(1-\omega_{1,2}^2)^{1/2}, \quad
   \alpha=\frac{\delta_2 (1+V_2)}{\delta_1 (1+V_1)}, \nonumber \\
   \beta=\alpha-1/\alpha, \quad
   \varepsilon=2\alpha-\beta+2(\omega_1\omega_2 -\eta_1 \eta_2 ),
   \quad  \tau=2(\omega_1 \eta_2 -\eta_1 \omega_2 ),\,\, \nonumber \\
   \varphi=2\alpha-\beta-2(\omega_1\omega_2 +\eta_1 \eta_2 ),
   \,\,\,
   \kappa=2(\omega_1 \eta_2 +\eta_1 \omega_2 ). \,\,\,\,\,\,\,\,\,\,\,\,
\end{align}
\end{widetext}
Here $x_{1,2}$ and $\Delta_{1,2}$ define initial positions and
initial phases of the two breathers, respectively.

It is possible to derive the separatrix solution from Eq.~(\ref{BB})
in the limits $V_{1,2}\rightarrow 0$ and $\omega_{1,2}\rightarrow 0$
but the derivation is tedious and for $N>3$ we calculate the maximal
energy density numerically considering collisions of slow kinks or
slow, low-frequency breathers. Parameters of the colliding solitons
are chosen to achieve collision of all $N$ subkinks at one point.
Note that the collisions of two slow, low-frequency breathers were
analyzed earlier in the study of fractal soliton collisions and the
possibility for all four subkinks to collide at one point was
demonstrated in Ref.~\cite{PREfractal}.

\begin{figure}
\includegraphics[width=9cm]{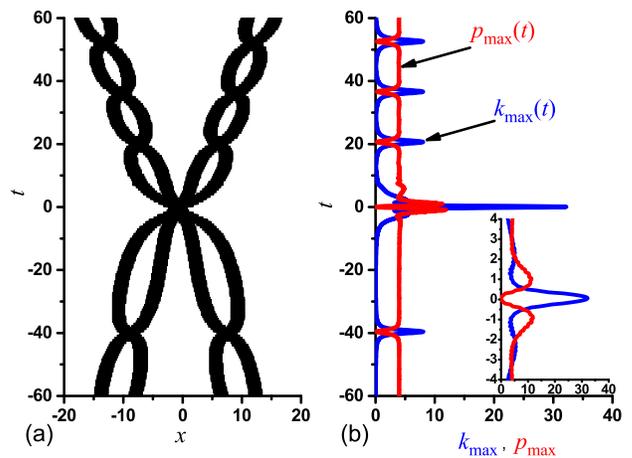}
\caption{The result of the numerical simulation of the collision
of two breathers (four subkinks). Initial conditions are set
with the help of Eq.~(\ref{BB}) with $V_{1}=-V_{2}=0.1$, $\omega_{1,2}=0.1$,
$\Delta_1=0$ and $\Delta_2=\pi$, and $x_{2}-x_{1}=4.012$. (a) the
soliton cores shown by
the regions where $e(x,t)>2$. (b) Maximal --over the
spatial coordinate $x$--
kinetic (blue) and potential (red) energy densities as functions
of time. The inset shows the curves near $t=0$. At the collision point
potential energy density is practically zero, while the kinetic energy
density increases up to nearly 32.} \label{fig6}
\end{figure}

Equations (\ref{SGEdiscrete}) are integrated numerically for $h = 0.1$,
$h = 0.05$ and $\tau = 0.005$. Initial conditions are set
with the help of Eq.~(\ref{BB}). For simplicity, the
collision of symmetric slow and low-frequency breathers is considered by setting
$V_{1}=-V_{2}=0.1$, $\omega_{1,2}=0.1$, $\Delta_1=0$ and $\Delta_2=\pi$.
To achieve the collision of all four subkinks at one point one should choose a proper initial
distance between the breathers. In a series of numerical runs it is
found that $x_{2}-x_{1}=4.012$ gives the desired result presented
in Fig.~\ref{fig6}.

It can be seen in Fig.~\ref{fig6} that at the point of collision of the four
subkinks the potential energy density is almost zero while the kinetic energy
density shows a peak with a height nearly equal to 32. More precisely, for
$h=0.1$ the largest energy density we could obtain by varying the parameter
$x_{2}-x_{1}$ was 32.21, while for $h=0.05$ it was 32.05. With decreasing $h$
the accuracy of simulation increases. We thus conclude that the total energy
density at the collision point is
\begin{equation}\label{K4max}
   e^{(4)}_{\max}=k^{(4)}_{\max}\approx 32.
\end{equation}

Note that after the collision breathers have frequencies and velocities
different from the initial values. This is due to the (weak but
still nontrivial in this collision phenomenon) effect of
discreteness, which breaks the integrability of the model. For more
details on the inelasticity of near-separatrix multi-soliton collisions
in weakly perturbed SGE see Refs. \cite{Mirosh,PREfractal,PREcollisions}.

\subsection{Case $N=5$} \label{Sec:N5}

\begin{figure}
\includegraphics[width=9cm]{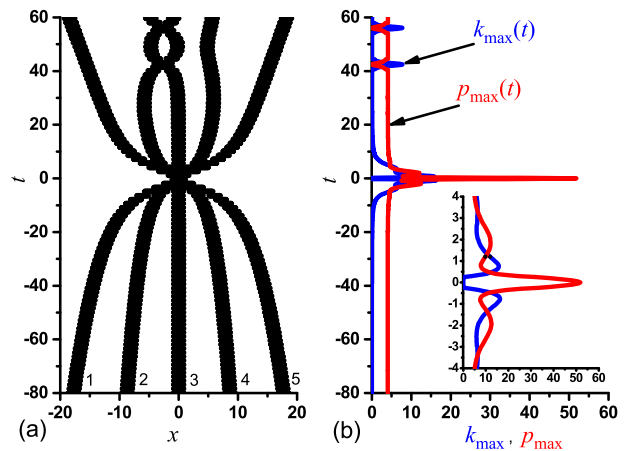}
\caption{Collision of five kinks/antikinks at one point. The choice
of initial conditions is described in the text. As before, in (a) the regions
of the $(x,t)$ plane with energy density $e(x,t)>2$ are shown. In (b)
the maximal, over $x$, kinetic and potential energy densities are shown as
functions of time by the blue and red lines, respectively. At the
collision point the potential energy density features the maximum of about
approximately 52,
while the kinetic energy density is almost zero. The inset shows the details of the curves
near $t=0$. Note that the solitons are numbered in (a) before the collision.
Odd quasi-particles are kinks and even ones are antikinks.
}\label{fig7}
\end{figure}

Here we set initial conditions using the individual kink (antikink) solution
of Eq.~(\ref{Kink}) [rather than an extremely cumbersome 5-soliton
solution]. As shown in Fig.~\ref{fig7}(a), the initial positions and
velocities of the five solitons are chosen such that initially they do
not overlap and so that they
collide at one point. As it was already mentioned, each
soliton should attract its nearest neighbors and thus, it should have the
topological charge opposite to that of its neighbors. In our case solitons 1, 3,
and 5 are kinks and 2 and 4 are antikinks. The kink 3 is located at the
origin and it is at rest, $x_3=0$ and $V_3=0$. The antikinks 2 and 4 have
velocities $V_2=-V_4=0.025$ and initial positions $x_2=-x_4=-12.0$. By symmetry
the solitons 2, 3, and 4 collide at one point. For the kinks 1 and 5 we take
two times larger velocities $V_1=-V_5=0.05$ and choose their initial coordinates
to achieve the collision of five solitons at one point. This happens for
$x_1=-x_5=-24.376549$. Although the exactly coincident collision
doesn't happen for exactly double initial distances (from the origin)
for double initial velocities, the latter is a reasonable rule of
thumb for preparing the initial conditions of the multi-soliton
configuration; a slight subsequent refinement may then be needed
(such as the slight displacement of the outer kinks from $x_1=-x_5=24$
to $x_1=-x_5=-24.376549$).

As it can be seen from Fig.~\ref{fig7}(b), when the five solitons collide,
the maximal
kinetic energy density is close to zero, while the
maximal potential energy density
is 50.93 for $h=0.1$ and 51.85 for $h=0.05$. We conclude that
\begin{equation}\label{K5max}
   e^{(5)}_{\max}=p^{(5)}_{\max}\approx 52.
\end{equation}
A relevant additional remark here is that the significant role of
weak asymmetries (in the preparation of our initial condition)
can be observed to be exacerbated in the outcome of the
collisional dynamics of Fig.~\ref{fig7}. In particular,
the figure showcases a visibly asymmetric result of the dynamics
featuring, in addition to two outer nearly symmetric
kinks, a breather (involving the anti-kink of soliton 2 and
the kink of soliton 3) and a ``stray'' kink (the antikink of soliton 4).
Once again here, the non-integrability of the underlying numerical
scheme is deemed to be responsible for the observed asymmetry,
although the energy density accumulation at $x=t=0$ is expected
to persist even for an integrable discretization.

\subsection{Case $N=6$} \label{Sec:N6}

\begin{figure}
\includegraphics[width=9cm]{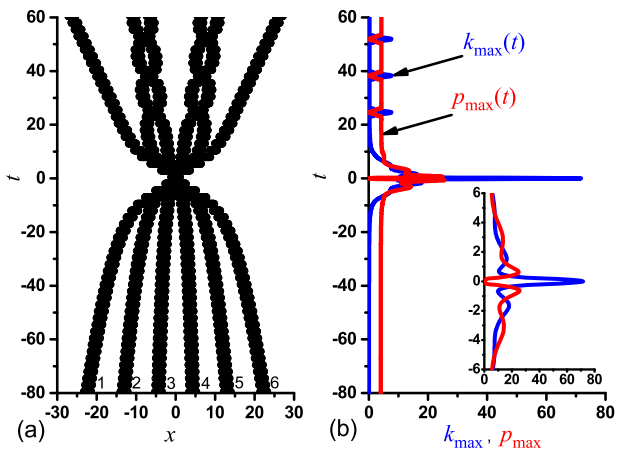}
\caption{Same as in Fig.~\ref{fig7} but for the six-soliton collision.
Initial conditions ensure the collision of all six kinks/antikinks
at one point (see the text for the details). When the kinetic energy density
reaches the maximal value of about 72, the potential energy density
is almost zero.} \label{fig8}
\end{figure}

Referring to Fig.~\ref{fig8}(a), note that the solitons 1, 3, and 5 are
kinks and 2, 4, and 6 are antikinks. Initial soliton positions and
velocities to achieve their collision at one point are:
$x_1=-x_6=-34.90395$, $V_1=-V_6=0.1$, $x_2=-x_5=19.37864$, $V_2=-V_5=0.05$,
$x_3=-x_4=-7$, $V_3=-V_4=0.025$. Once again, the velocities have
been selected using factors of $2$, while the positions have been
refined (from the corresponding factors of $2$) to ensure that the
collision occurs for all solitons at the same point.

From Fig.~\ref{fig8}(b) it is clear that at the collision point the
maximal, over $x$, potential energy density is nearly zero while the maximal
kinetic energy density reaches its highest attainable
value. The height of the maximum
is 72.62 for $h=0.1$ and 72.08 for $h=0.05$. Thus,
\begin{equation}\label{K6max}
   e^{(6)}_{\max}=k^{(6)}_{\max}\approx 72.
\end{equation}
Here, solitons 2 and 3, as well as 4 and 5 merge in the symmetric
aftermath of the collision into breather states (a feature that once
again would be avoided in the realm of fully integrable dynamics).

\subsection{Case $N=7$} \label{Sec:N7}

\begin{figure}
\includegraphics[width=9cm]{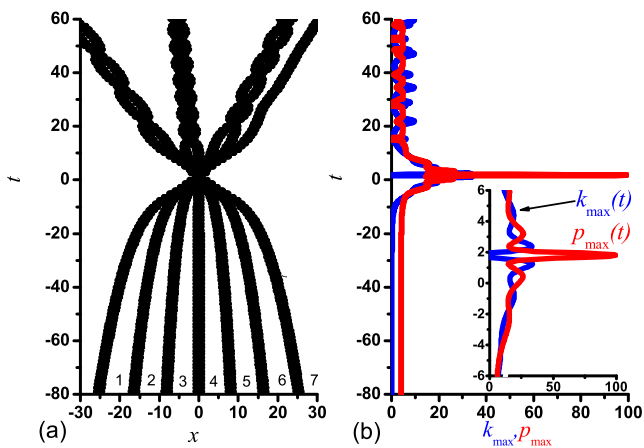}
\caption{Same as in Fig.~\ref{fig7} but for the seven-soliton collision.
All seven kinks/antikinks collide at one point due to proper choice of
the initial conditions (see the text for the details). When
the potential energy density
reaches the maximal value of about 100, the kinetic energy density
is almost zero.}
\label{fig9}
\end{figure}

In the initial configuration, odd solitons in Fig.~\ref{fig9}(a) are
the kinks and even are the antikinks. They collide at one point provided
that their initial coordinates and velocities are chosen as follows:
$x_1=-x_7=-39.541867403$, $V_1=-V_7=0.1$, $x_2=-x_6=-24.29923$,
$V_2=-V_6=0.05$, $x_3=-x_5=-12$, $V_3=-V_5=0.025$, and $x_4=0$, $V_4=0$.
Looking at Fig.~\ref{fig9}(b) we note that at the collision point the
maximal over $x$ kinetic energy density is extremely small, while the
maximal potential energy density features a maximum of 94.90 for $h=0.1$
and 99.56 for $h=0.05$. It can then be stated that
\begin{equation}\label{K7max}
   e^{(7)}_{\max}=p^{(7)}_{\max}\approx 100.
\end{equation}

The case of Fig.~\ref{fig9} is once again one of a pronounced asymmetric
outcome, as we have generally observed odd $N$ cases to be
(cf. Fig.~\ref{fig7}). Three breathers are observed to form
(solitons 2-3, 4-5, and 6-7), while the 1st soliton moves to the right
in an isolated trajectory.

\section {Conclusions \& Future Challenges} \label{Sec:Conclusions}

In this work, we have
provided a systematic calculation of
the maximal energy density in the collision of
$N$ slow kinks/antikinks (with $N \leq 7$)
in the integrable sine-Gordon model.
Our findings are collected in Table~\ref{cnls_numer_summary}. The first
line gives the number of colliding solitons, $N$. The second line gives
the {\em exact} values of the maximal energy density that can be achieved in
the collision of $N$ kinks/antikinks.
These results are available for $N\le 3$ (see Secs. \ref{Sec:N1}, \ref{Sec:N2},
and \ref{Sec:N3}). For larger $N$ the results were obtained numerically and
they are presented in the third and fourth lines of Table~\ref{cnls_numer_summary} for
$h=0.1$ and $h=0.05$, respectively. In numerical simulations the kink/antikink
velocities are small (no greater than 0.1) but not equal to zero at
$t\rightarrow \pm\infty$. For decreasing $h$ and decreasing initial
velocities of the colliding kinks/antikinks the numerical results converge
to the integer numbers shown in the last two lines of Table~\ref{cnls_numer_summary}.

\begin{table}[pht]
\caption{Summary on maximal energy density in collision of $N$ solitons for $N\le 7$.} \centering
\begin{tabular}{ | l | c | c | c |c | c | c | c | r|  }
\hline
$N$        & 0 & $1$ & $2$ & $3$  & $4$     & $5$     & $6$     & $7$     \\
Exact      & 0 & $4$ & $8$ & $20$ & $-$     & $-$     & $-$     & $-$     \\
$h=0.1$    &   & $$  & $$  & $$   & $32.21$ & $50.93$ & $72.62$ & $94.90$ \\
$h=0.05$   &   & $$  & $$  & $$   & $32.05$ & $51.85$ & $72.08$ & $99.56$ \\
  \multicolumn{9}{|l|}{}                             \\
$2N^2$     &$0$& $-$ & $8$ & $-$  & $32$    & $-$     & $72$    &  $-$    \\
$2(N^2+1)$ &$-$& $4$ & $-$ & $20$ & $-$     & $52$    & $-$     & $100$   \\ \hline
\end{tabular}
\label{cnls_numer_summary}
\end{table}

The results can be summarized as follows. The maximal energy density
that can be achieved in collision of $N$ slow kinks/antikinks in SGE is
found to be equal to
\begin{eqnarray}\label{fit}
   e_{\max}^{(N)}\approx 2N^2 \quad &&{\rm for \,\, even }\,\, N, \nonumber \\
   e_{\max}^{(N)}\approx 2(N^2+1) \quad &&{\rm for \,\, odd }\,\, N.
\end{eqnarray}

When an even number of slow kinks/antikinks collides at one point,
the kinetic energy density reaches a maximal value $2N^2$,
while the maximal potential energy density is nearly equal to zero.
On the contrary, when an
odd number of slow kinks/antikinks collides at one point,
the potential energy density has maximal value $2 (N^2+1)$,
while the maximal kinetic energy density is almost zero.

These maximal energy density values can be achieved when all $N$ kinks/antikinks
collide at one point. This happens when the kinks and antikinks approach
the collision point alternatively (i.e., no two adjacent solitons
are of the same type). Arranged in this way, each soliton has nearest
neighbors of the opposite topological charge. Such solitons attract each other
and their cores can merge producing a controllably
high energy density spot, as we have demonstrated herein.

According to Eq.~(\ref{fit}), the maximal energy density in the sine-Gordon
field that can be realized in $N$-soliton collisions increases quadratically
with $N$. At the same time, total energy of $N$ standing kinks is equal to
$8N$ and thus, is proportional to $N$. Naturally, this does not lead to
a contradiction since the very high energy density is accumulated
at a very narrow region near $x=0$, and hence when integrated over
space, still preserves the total energy of $8 N$. Furthermore, this
very high concentration of energy density for a very short time
interval (around $t=0$) is reminiscent of rogue events in other
models (such as the nonlinear Schr{\"o}dinger equation and variants
thereof, with their Peregrine soliton and related solutions)~\cite{pelin,yan}.
However, to the best of our knowledge, no explicit
rogue waveforms have been identified yet in
such models. Hence, our identification of controllably
large energy densities in the SGE model is, arguably, the first
example of such a rogue event in this setting.

Having the results of this work in mind,
one can expect that in the soliton gas model \cite{Baryakhtar,Baryakhtar1}
unlimited energy density can be achieved. Of course, the probability of
collision of $N$ alternating kinks and antikinks decreases rapidly with
increasing $N$ (and even then, the probability of their
concurrent collision is very low),
but such rare events can have important consequences, when they
do arise.

As for the open problems, it is important to calculate the maximal
energy density that can be achieved in multi-soliton collisions in other integrable and non-integrable
systems of different dimensionality. For example, one can examine similar
issues and design such collisions in other Klein-Gordon field theoretic
models (e.g. in the $\phi^4$ or $\phi^6$ models~\cite{Backlund1,Gani}), as well as in the
one-dimensional, self-defocusing nonlinear Schr{\"o}dinger equation.
It would be particularly interesting to explore if the relevant
phenomenology persists therein. It would also be particularly interesting
to explore to attempt to prove the asymptotic statements inferred
herein; although perhaps a direct approach towards this starting
from a multi-soliton solution could be very cumbersome, perhaps
a reverse approach, initializing the system with a suitably large,
and highly localized energy density at a point and utilizing the
inverse scattering transform to establish that this waveform
will split into $N$ soliton solutions may be more tractable.

\section*{Acknowledgments}

D.S. thanks the financial support of the Institute for Metals
Superplasticity Problems, Ufa, Russia. S.V.D. thanks financial
support provided by the Russian Science Foundation grant 14-13-00982.
P.G.K. acknowledges
support from the US National Science Foundation under grants
DMS-1312856, from FP7-People under grant
IRSES-605096 from the Binational (US-Israel) Science Foundation
through grant 2010239, and from the US-AFOSR under grant
FA9550-12-10332.


\begin{thebibliography}{99}

\bibitem{Scott} A.C. Scott, Am. J. Phys. 37, 52 (1969).

\bibitem{BookSGE} J. Cuevas-Maraver, P.G. Kevrekidis, and F. Williams, eds.,
The sine-Gordon Model and its Applications. From Pendula and Josephson
Junctions to Gravity and High Energy Physics, Springer, Berlin, 2014.

\bibitem{Eisenhart}
L.P. Eisenhart, A treatise on the differential geometry of curves and surfaces,
Ginn and Co., Boston, 1909.

\bibitem{Watanabe}
S. Watanabe, H.S.J. van der Zant, S.H. Strogatz, T.P. Orlando, Physica D {\bf 97}, 429 (1996).

\bibitem{Perring}
J.K. Perring, T.H.R. Skyrme, Nucl. Phys. {\bf 31}, 550 (1962).

\bibitem{Coleman}
S. Coleman, Phys. Rev. D {\bf 11}, 2088 (1975).

\bibitem{BraunKivshar} O.M. Braun, Yu.S. Kivshar, \textit{The Frenkel-Kontorova Model:
Concepts, Methods, and Applications} (Springer, Berlin, 2004).

\bibitem{Ekomasov}
E.G. Ekomasov, R.R. Murtazin, O.B. Bogomazova, A.M. Gumerov, J. Magn. Magn. Mater. {\bf 339}, 133 (2013).

\bibitem{Ferro1}
S.V. Dmitriev, K. Abe, T. Shigenari, Physica D {\bf 147}, 122 (2000).

\bibitem{Ferro2}
S.V. Dmitriev, K. Abe, T. Shigenari, J. Phys. Soc. Jpn {\bf 65}, 3938 (1996).

\bibitem{Drazin}
P.G. Drazin, Solitons, in: London Mathematical Society Lecture Note Series, Vol. 85,
Cambridge University Press, Cambridge, 1983.

\bibitem{Barone}
A. Barone, F. Esposito, C.J. Magee, A.C. Scott, Riv. Nuovo Cimento {\bf 1}, 227 (1971).

\bibitem{KivshMal} Yu. S. Kivshar and B.A. Malomed
Rev. Mod. Phys. {\bf 61}, 763 (1989).

\bibitem{weinbirn} B. Birnir, H.P. McKean and A. Weinstein,
Comm. Pure Appl. Math. {\bf 47}, 1043 (1994).

\bibitem{Backlund1}
R.K. Dodd, J.C. Eilbeck, J.D. Gibbon, H.C. Morries, {\it Solitons
and Nonlinear Wave Equations} (Academic Press, London, 1982).

\bibitem{Backlund2} A.P. Fordy, A historical introduction to solitons
and B\"acklund transformations. In A.P. Fordy, J.C. Wood, eds,
{\it Harmonic Maps and Integrable Systems}, (Vieweg, Wiesbaden,
1994), PP. 7-28.

\bibitem{Hirota}
R. Hirota, J. Phys. Soc. Jpn {\bf 33}, 1459 (1972).

\bibitem{Pol}
L.A. Ferreira, B. Piette, and W.J. Zakrzewski,
Phys. Rev. E {\bf 77}, 036613 (2008).


\bibitem{giamar} L. Benfatto, C. Castellani, and T. Giamarchi, Phys. Rev.
Lett. {\bf 99}, 207002 (2007).

\bibitem{takacs} Z. Bajnok, L. Palla, and G. Takacs, Nucl. Phys. B
{\bf 644}, 509
(2002); Z. Bajnok, C. Dunning, L. Palla, G. Takacs, and F. Wagner,
Nucl. Phys. B {\bf 679}, 521 (2004).

\bibitem{malda} D. M. Hofman and J. M. Maldacena, J. Phys. A
{\bf 39}, 13095
(2006).

\bibitem{hindmar} P. Salmi, M. Hindmarsh,
Phys. Rev. D {\bf 85}, 085033 (2012).

\bibitem{skyr} S.B. Gudnason, M. Nitta,
Phys. Rev. D {\bf 90}, 085007 (2014).

\bibitem{pelin} A. Slunyaev, I. Didenkulova, E. Pelinovsky,
Cont. Phys. {\bf 52}, 571 (2011); see also:
C. Kharif, E. Pelinovsky and A. Slunyaev, {\it Rogue Waves in
the Ocean}, Springer-Verlag (Berlin, 2009).

\bibitem{yan} Z. Yan,
J. Phys. Conf. Ser. {\bf 400}, 012084 (2012).

\bibitem{Legrand} O. Legrand, G. Reinisch, Phys. Lett. A {\bf 35}, 3522 (1987).

\bibitem{Legrand1} O. Legrand, Phys. Lett. A {\bf 36}, 5068 (1987).

\bibitem{CaputoFlytzanis} J.G. Caputo, N. Flytzanis, Phys. Rev. A {\bf 44}, 6219 (1991).


\bibitem{Danial}
D. Saadatmand, S.V. Dmitriev, D.I. Borisov, P.G. Kevrekidis, Phys. Rev. E {\bf 90}, 052902 (2014).

\bibitem{PREcollisions} S.V. Dmitriev, P.G. Kevrekidis, Yu.S.
Kivshar, Phys. Rev. E {\bf 78}, 046604 (2008).

\bibitem{Mirosh}
A.E. Miroshnichenko, S.V. Dmitriev, A.A. Vasiliev, T. Shigenari,
Nonlinearity {\bf 13}, 837 (2000).

\bibitem{PREfractal} S.V. Dmitriev, Yu.S. Kivshar, T.
Shigenari, Phys. Rev. E {\bf 64}, 056613 (2001).

\bibitem{Baryakhtar} I.V. Baryakhtar, V.G. Baryakhtar, E.N. Economou, Phys. Rev. E {\bf 60}, 6645 (1999).

\bibitem{Baryakhtar1} I.V. Baryakhtar, V.G. Baryakhtar, E.N. Economou, Phys. Lett. A {\bf 207}, 67 (1995).

\bibitem{Gani} V.A. Gani, A.E. Kudryavtsev, and M.A. Lizunova,
Phys. Rev. D {\bf 89}, 125009 (2014).

\end{thebibliography}
\end{document}